# The shock-heated atmosphere of an asymptotic giant branch star resolved by ALMA

**Authors:** Wouter Vlemmings[1,*], Theo Khouri[1], Eamon O'Gorman[2], Elvire De Beck[1], Elizabeth Humphreys[3], Boy Lankhaar[1], Matthias Maercker[1], Hans Olofsson[1], Sofia Ramstedt[4], Daniel Tafoya[1], Aki Takigawa[5]

**Affiliations:**

[1]Department of Space, Earth and Environment, Chalmers University of Technology, Onsala Space Observatory, 439 92, Onsala, Sweden.

[2]Dublin Institute for Advanced Studies, 31 Fitzwilliam Place, Dublin 2, Ireland.

[3]ESO Karl-Schwarzschild-Str. 2, 85748, Garching, Germany.

[4]Department of Physics and Astronomy, Uppsala University, Box 516, 751 20, Uppsala, Sweden.

[5]The Hakubi Center for Advanced Research/Division of Earth and Planetary Sciences, Graduate School of Science, Kyoto University, Kitashirakawa-Oiwakecho, Sakyo, Kyoto 606-8502, Japan

**Our current understanding of the chemistry and mass-loss processes in solar-like stars at the end of their evolution depends critically on the description of convection, pulsations and shocks in the extended stellar atmosphere *(1)*. Three-dimensional hydrodynamical stellar atmosphere models provide observational predictions *(2)*, but so far the resolution to constrain the complex temperature and velocity structures seen in the models has been lacking. Here we present sub-millimeter continuum and line observations that resolve the atmosphere of the asymptotic giant branch star W Hya. We show that hot gas with chromospheric characteristics exists around the star. Its filling factor is shown to be small. The existence of such gas requires shocks with a cooling time larger than commonly assumed. A shocked hot layer will be an important ingredient in the models of stellar convection, pulsation and chemistry that underlie our current understanding of the late stages of stellar evolution.**

Asymptotic giant branch (AGB) stars are among the most important sources of enrichment of the Galactic interstellar medium (ISM). Molecules and dust formed in the warm extended atmospheres and the cool and dense circumstellar envelopes (CSEs) around AGB stars are injected into the ISM by a stellar wind that has overcome stellar gravity *(1)*. It is generally assumed that the stellar wind is driven by radiation pressure on dust that forms at a few stellar radii, where the temperature in the CSE has dropped so that dust condensation can occur *(3)*. In order for the gas in the extended stellar atmosphere to reach the dust formation region, the most recent AGB mass-loss models typically invoke stellar pulsations and convective motions *(2, 4-6)*.

Both convective motions and pulsations induce outward moving shocks that critically affect the upper layers of the AGB atmosphere where the stellar mass loss is determined. The propagation of shocks also strongly affects the chemistry in the stellar atmosphere *(7-9)*. In early AGB atmosphere models, the outward propagation of strong shocks is responsible for the creation of a chromosphere *(6, 10)*, from which ultraviolet line and continuum emissions originate. Such emissions have been observed from AGB stars *(11, 12)*. However, the observations of molecules and dust close to the star are not consistent with the extended chromosphere produced by the models. Observations have so far not been able to resolve this ambiguity. High angular resolution images of the stellar disks of AGB stars have revealed asymmetries of which the source is not yet clear, but convective motions are believed to play a role *(13-16)*. Since at most wavelengths, the observations are probing distinct molecular opacity sources *(16)*, or averages over the stellar disk *(17)*, the dynamics and temperature structures in the atmosphere closest to the stellar photosphere have not yet been observed in detail.

We present observations of the AGB star W Hya that reveal evidence for the presence of shocks and map the distribution of molecular gas in the extended atmosphere. W Hya, with a regular pulsation period of 361 days, and a mass-loss rate of $\sim 1.3 \times 10^{-7}$ $M_\odot$ yr$^{-1}$, has been observed in detail across a large optical to radio wavelength range *(15, 18-20)*. It is estimated to have a photospheric radius $R_\star \sim 20$ mas (2 au, for an adopted distance of 98 parsec *(21)*) *(19, 20)* and is surrounded by an irregular dust formation region at approximately 1.7 to 2.5 $R_\star$ *(15)*. W Hya emits strongly in the ultraviolet *(12)*, which likely indicates it has a hot chromosphere, although the stellar photosphere has a temperature of only 2500 K (*18*). At radio frequencies (43 GHz), marginally resolved observations show that it is surrounded by an extended emission region of 3.4x2.3 $R_\star$ *(22)*. This extended region has been named the radio photosphere. In the currently accepted model, emission of the radio-photosphere is free-free emission originating from free electrons interacting with neutral hydrogen atoms and molecules. The free electrons are provided by the ionization of atoms with a low ionization potential *(23)*. At different frequencies, the free-free emission becomes optically thick at different heights in the atmosphere. Hence, the brightness temperature of the radio emission directly reflects the local gas temperature. The radio observations, which are consistent with a steady, outwardly decreasing temperature, thus appear to show no indication of an extended hot chromospheric layer. The solution to the apparent discrepancy between the ultraviolet and the radio observations could be that the hot chromospheric gas has a very low filling factor, as previously suggested to explain observations of the supergiant Betelgeuse at radio wavelenghts *(24,25)*.

With the long baselines of the Atacama Large (sub-)Millimeter Array (ALMA), it is now possible to resolve the stellar disk of the most nearby AGB stars at frequencies that probe deep into the extended stellar atmosphere. Using 8 km baselines, data on the AGB star W Hya were obtained around 338 GHz in December 2015. At that point in time, W Hya was in stellar pulsation phase $\phi$~0.3. Details of the observations are presented in the Method section. The resulting brightness temperature image of the stellar disk is shown in Fig.1. We have fitted the star, which has a flux density of 470 ± 48 mJy, with a uniform elliptical disk. The disk has an average brightness temperature of 2495 ± 255 K. It has a major axis of 56.5 ± 0.1 mas (5.53 ± 0.01 au, 2.79 ± 0.01 $R_\star$), a minor axis of 51.0 ± 0.1 mas (5.00 ± 0.01 au, 2.50 ± 0.01 $R_\star$), and a position angle of 65.7 ± 0.3 degrees. Assuming optically thick free-free emission, the brightness temperature is equal to the gas temperature. The average gas temperature is thus similar to the temperature of the stellar photosphere, even though the average disk radius is approximately 30% larger than that of the photosphere at comparable phase. The observed disk radius (1.3 $R_\star$) implies that the submillimeter wavelength observations probe closer to the photosphere than observations at most other wavelengths. Dust emission, molecules and atmostpheric absorption affect the measurements in significant parts of the near-infrared and optical regimes *(20)*.

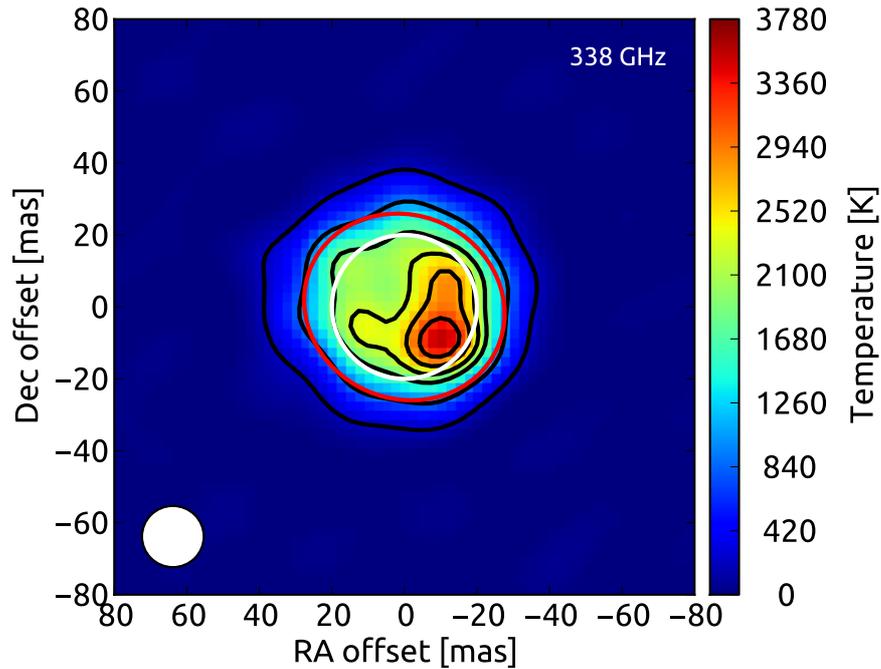

*Fig. 1*. Brightness temperature map of the surface of the AGB star W Hya as observed with ALMA at 338 GHz. Contours are drawn at 227, 907, 1814, 2268, 2722, and 3194 K. The rms noise in the image is 18 K and the peak is 3560 K. The image is centered at RA(J2000) = $13^h49^m01.939^s$ and Dec(J2000) = $-28^d22^m04.484^s$. The map was produced using uniform weighting of the interferometric visibilities and was restored with a circular beam (shown in the bottom left) of 17 mas (~1.7 au), corresponding to the minor axis of the 17x24 mas dirty beam. The red ellipse indicates the size of the fitted, uniform, stellar disk (51x56.5 mas, PA 65.7 degrees) with a brightness temperature of 2495 ± 125 K. The white circle denotes the stellar photosphere estimated from near-infrared observations (19). The effective temperature of the photosphere is ~2500 K (18).

It is immediately apparent that the stellar disk is far from uniform. We can identify regions both hotter and cooler than the photospheric temperature. Further asymmetries in the limbs of the disk become apparent when the average disk is subtracted from the interferometric visibilities and a residual image is generated (Fig.2). The brightest of the hotspots, in the southwest quadrant of W Hya, is unresolved in our observations with a beam of 17 mas (1.7 au, 0.9 $R_\star$). Because of the high signal-to-noise detection, the size of the hotspot can conservatively be constrained to be less than 2x3 mas (see Method section). This implies that its area is less than 0.05 au$^2$, corresponding to an area filling factor <0.2%. Assuming a Gaussian structure, this means that the true brightness temperature of the spot is > 5.3x10$^4$ K, consistent with the temperature expected in a strong shock *(7, 26)*. Overall, almost half of the stellar disk, at 17 mas resolution, has a temperature elevated above the photospheric temperature. This apparent high temperature could be caused by a combination of the several strong compact and unresolved hotspots that are directly detected and/or additional weaker hotspots and more extended gas with a $T_{gas}$~2700 K. The rest of the disk, and specifically the northeast quadrant, has a similar or slightly below average temperature.

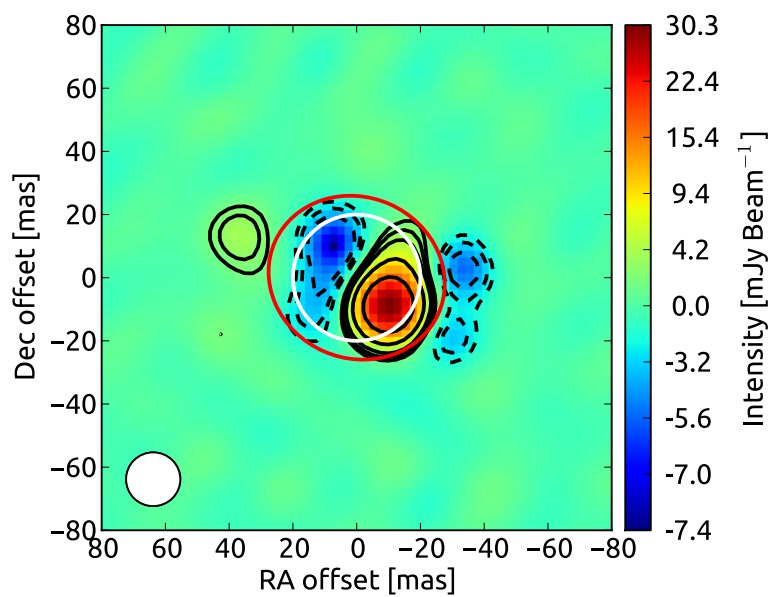

***Fig. 2.*** *The residual map after subtracting the fitted uniform elliptical disk from the*

*continuum image of W Hya. Contours are drawn at -9, -6, -4, 4, 6, 9, 12, and 24 times the image rms noise of 0.45 mJy Beam$^{-1}$. The beam size and ellipses are as described with Fig. 1. Significant deviations from a uniform disk are obvious. The brightest spot is unresolved in our observations, indicating a brightness temperature $T_b>5.3x10^4$ K.*

In addition to the stellar continuum emission, the observations also covered the pure-rotational CO transition, *J=3-2,* in the first vibrationally excited state, *v=1,* at 342.647 GHz. This transition, with an upper level at ~3120 K, is a unique probe of the warm gas in the extended AGB atmosphere *(27)*. The line was detected both in absorption towards the star and in emission around the star, as indicated in the velocity channel maps shown in Supplementary Figure 1. The high spatial resolution allowed us to determine kinematically distinct infall and outflow components in the extended atmosphere across the star. In Fig. 3, we show three spectra, taken towards the eastern (cool) and western (hot) hemisphere of the star as well as in an annulus around the star. Using these spectra we constrained the gas velocity, mass and temperature.

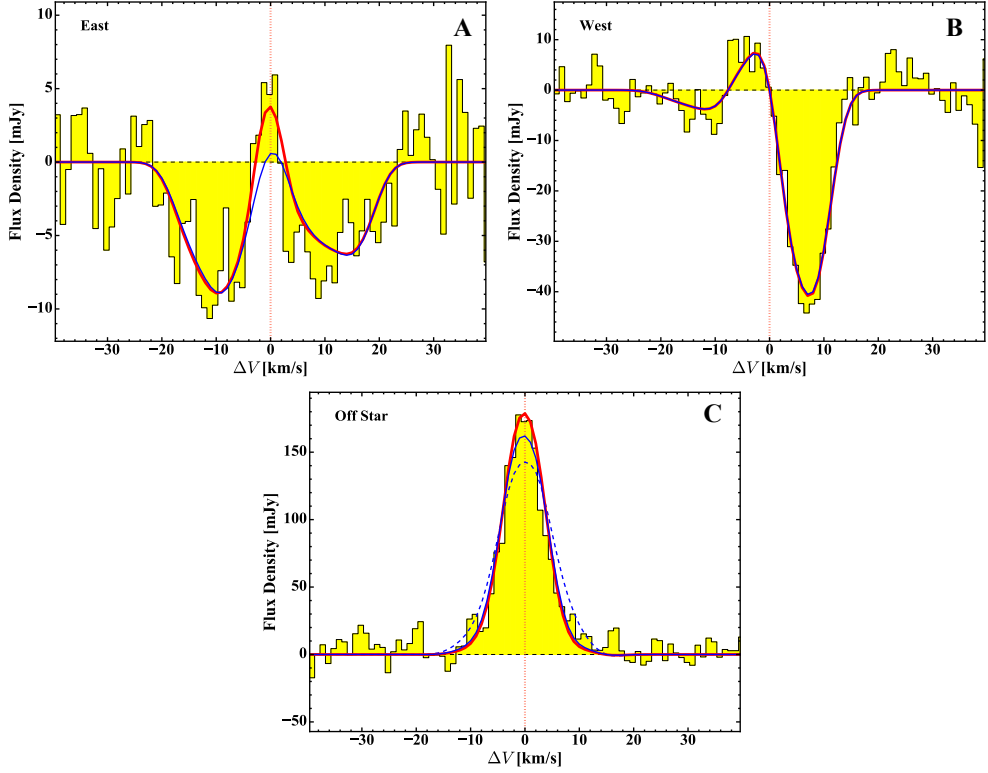

**Fig. 3.** *Spectra of the CO J=3-2, v=1 transition extracted towards the eastern (A) and western (B) stellar hemisphere as well as in an annulus between 2.9 and 5 au off the star (C). The spectra have a resolution of ~1 km s$^{-1}$. The spectra are presented relative to the stellar velocity of 39.2 km s$^{-1}$. The thick red lines gives our model results when including a warm and a cooler molecular gas layer of ~2900 K and ~850-900 K respectively that best describe the observations. The fits are described in the Method section. Further parameters of the models are given in Table. 1. The thin blue line indicates our model results without the warm molecular gas layer. In panel C, the lines are an average of the two hemispheres except for the dashed blue line, which indicates specifically the model of the eastern hemisphere without a warm layer. A similar line for the western hemisphere is omitted since it would be nearly identical to the blue line. The models towards the hemispheres are constrained by the emission in the surrounding annulus. Especially for the eastern hemisphere, models without a warm layer are unable to reproduce the observations. For the western hemisphere, the models with and without the warm layer are nearly identical.*

| | | | Temperature [K] | Mass [$M_\odot$] | Velocity (Range) [km s$^{-1}$] |
|---|---|---|---|---|---|
| East | 'Warm' | post-shock | $2900^{+400}_{-500}$ | $(1.0^{+2}_{-0.5}) \times 10^{-5}$ | $(+1^{+2}_{-1})$ |
| | 'Cool' | Infall | $900 \pm 100$ | $(4.0 \pm 1.0) \times 10^{-5}$ | $(-21 \pm 3)$ to $0$ |
| | 'Cool' | Outflow | $900 \pm 100$ | $(2.7 \pm 0.8) \times 10^{-5}$ | $0$ to $(+20 \pm 3)$ |
| West | 'Warm' | post-shock | $2900^{+400}_{-500}$ | $(1.0^{+2}_{-0.5}) \times 10^{-5}$ | Unconstrained |
| | 'Cool' | Infall | $850 \pm 50$ | $(7.0 \pm 1.0) \times 10^{-5}$ | $(-13 \pm 1)$ to $(-2 \pm 1)$ |
| | 'Cool' | Outflow | $850 \pm 50$ | $(5.0 \pm 1.0) \times 10^{-6}$ | $(+10 \pm 2)$ to $(+20 \pm 3)$ |

***Table. 1.*** *Mass, average temperature, and velocity (range) for the molecular gas components that best match the observed spectra of the CO J=3-2, v=1 transition shown in Fig.4. The eastern and western hemisphere models are determined as described in the Method section and are matched to the respective hemisphere and off-star annulus simultaneously. The error bars indicate formal one-sigma uncertainties. The velocities indicate outflow (positive) and infall (negative) with respect to a stellar velocity of 39.2 km s$^{-1}$. For the warm layer, we fit a single velocity instead of a range. The temperature of the warm layer is degenerate with the mass, with the lowest temperature limit corresponding to the largest mass and vice versa.*

The vibrationally excited CO spectra are best described by two components, a warm molecular gas layer close to the stellar surface we measured at 338 GHz, and cooler-gas infall and outflow components out to a radius of approximately 2.5 $R_\star$ (50 mas), encompassing the dust formation region. We find that the kinematics of the components above the cooler part of the stellar surface are markedly different from those towards the hotspot. The parameters of our models are given in Table 1, while details of the models are presented in the Method section. In our model the warm gas layer has $T_{gas}$ = 2900 K. The cool gas layers have an average temperature $T_{gas}$ ~ 850-900 K. The total mass in the visible warm layer, not hidden behind the optically thick continuum surface, is ~ $2 \times 10^{-5}$ $M_\odot$, with an uncertainty of a factor of two. The molecular gas mass in the cooler region between 2.6 and 5 au, encompassing the dust formation region, is $(1.4 \pm 0.4) \times 10^{-4}$ $M_\odot$. Hence, the mass ratio between warm and cool gas is between 0.06 and 0.4. Assuming at least an order of magnitude higher density in the warm gas region, as expected in the post-shock environment *(26)*, the volume-filling factor of the warm gas is <4%. In addition to the very hot (non-molecular) gas probed by the continuum emission, we thus find evidence for a significant amount of warm molecular gas in the immediate stellar environment of W Hya with a temperature above that of the photosphere. Several decades ago, similar layers were inferred from unresolved infrared observations *(28)*. Additional evidence for high excitation conditions in a warm layer is provided by the detection of an unidentified line in emission on top of the stellar continuum (Method section).

Towards the hotspot, the total gas content is dominated by an infall component. Our inferred warm layer only has a low outflow velocity relative to the star compared to the magnitude of the outflow and infall velocities of the cooler gas. Against the cooler part of the star, we more clearly see the warm layer. We also see fast outflowing and infalling gas over a velocity range of ~40 km s$^{-1}$. The highest outflow velocity that matches the spectra is ~20 km s$^{-1}$ which equal to the escape velocity from W Hya at 1.8 $R_\star$, assuming the star has a mass of ~1 $M_\odot$ *(29)*.

The total mass traced by the vibrationally excited CO is over three orders of magnitude more than the mass lost by W Hya in the course of one pulsation period. Thus, the gas will spend at least $10^3$ year in this region or even closer to the photosphere. Despite the low filling factor of the shocks, the fact that the shock and/or stellar activity heating timescales are significantly shorter implies most of the eventually ejected gas will have spent time in a shock-heated state. This will significantly affect the chemistry of the extended AGB atmosphere, indicating that non-equilibrium chemistry such as those in shock chemistry models *(7-9)* must be taken into account.

The detection of temperatures of the continuum that are significantly elevated compared to the stellar photosphere challenges most of the recent models used to describe AGB mass loss. Only the pulsation models where the shock energy is not immediately radiated away reach temperatures comparable to our measurements *(6)*. However, as these models typically predict a chromosphere extending significantly beyond the dust formation zone, the post-shock cooling time is strongly debated *(3, 10)* and in more recent models is assumed to be nearly instantaneous. Our observations now show that the cooling time cannot be instantaneous. The significant differences in kinematics across the stellar disk further indicate that global pulsation models are insufficient.

Recently, multi-dimensional models have focused on convection. In these models, the surface of an AGB star is made up of only very few large individual convective cells *(2)*. Our observations also reveal a mottled surface with at least one distinct hotspot. However, the estimated size of this spot is significantly less than expected for convective cells. The velocities we measure across the stellar disk are also not consistent with current models. At the radii where we trace the gas, the models predict infalling and outflowing motions with a relative velocity of less than 6 km s$^{-1}$, at least a factor of three less than what we observe. Improvements in resolution and in the treatment of shocks and pulsations might reconcile the convection models with the observations. Alternatively, the compact hotspot could be the result of solar-like magnetic activity. A weak, ~2 Gauss, magnetic field has been detected at the surface of one AGB star *(30)*, and observations of masering gas in the CSE of several more stars indicate strong magnetic fields could be present on the stellar surface *(31)*.

*Correspondence to: wouter.vlemmings@chalmers.se

**Acknowledgments:** Support for this work was provided by the European Research Council (ERC) through the ERC consolidator grant nr. 614264, and by the Swedish Research Council (VR). EDB further acknowledges support from the Swedish National Space Board (SNSB). We are also indebted to the staff of the Nordic ALMA ARC node, and in particular Ivan Marti-Vidal, for developments of the tools used in the data analysis and plotting. ALMA is a partnership of ESO (representing its member states), NSF (USA) and NINS (Japan), together with NRC (Canada), NSC and ASIAA (Taiwan), and KASI (Republic of Korea), in cooperation with the

Republic of Chile. The Joint ALMA Observatory is operated by ESO, AUI/NRAO and NAOJ.

**Author contributions:** W.V. reduced and analyzed the data, and wrote most of the manuscript. T.K. performed the radiative transfer modeling and wrote the modeling section of the Methods. E.D.B. and B.L. analyzed the unidentified line and provided the relevant text. A.K. obtained the ALMA data. All authors contributed with comments on the manuscript and data interpretation.

**Competing interests:** The authors declare no competing financial interests.

**Methods:**

1) **ALMA Observations**

The AGB star W Hya was observed with ALMA in three observing blocks spread over one week. The first observation was taken Nov 30$^{th}$ 2015, the second on Dec 3$^{rd}$ 2015, and the third on Dec 5$^{th}$ 2015. The observations spanned a total time of 4.25 hours, of which approximately 2.3 hours were spent on source. The observations were done using 47 antennas in an extended configuration, with baseline lengths between 18 and 8283 m. Three spectral windows were used in spectral line mode. Two spectral windows of 937.5 MHz and 960 channels were centered on 330.87 (spw0) and 342.87 GHz (spw1), and a window of 1.875 GHz and 3840 channels was centered on 344.97 GHz (spw2).

The data were reduced using CASA 4.7.0. We adopted the calibration approach of the observatory provided scripts. However, in addition to four antennas for which several time ranges were flagged in the script, we found that an additional antenna needed to be flagged due to high system temperatures. Flux and bandpass calibration were done on the quasar J1337-1257, for which a flux of 1.437 Jy at 339.57 GHz, and a spectral index of -0.77, were adopted. Based on flux calibration monitoring results of J1337-1257, the flux measurements of the phase calibrator (J1351-2912 with an adopted flux of 93, 82 and 78 mJy for spw0, spw1, and spw2), and uncertainties in the models for the solar system objects used in the calibrator monitoring, we conclude that the maximum absolute flux uncertainty is 10%.

After the script calibration, we performed two rounds of phase self-calibration and one round of phase and amplitude self-calibration on the stellar continuum. This improved the dynamic range from ~380 to ~2860. Using robust Briggs weighting in the CASA task *tclean*, we obtained a synthesized beam of 0.031x0.038" at position angle -84.3 deg. Using uniform weighting, we obtained a beam of 0.017x0.024" at a similar position angle at the expense of dynamic range. As the signal-to-noise is still high, we produced our final images using uniform weighting and a 17 mas restoring beam. We have confirmed that the beam selection does not affect the flux density in our spectra and the measured residuals. For the line imaging, we used a ~1 km s$^{-1}$ spectral resolution.

The brightness temperature image was created from the total intensity image using the formula for the brightness temperature of an elliptical Gaussian:

$T_b = \frac{1.36 \cdot 10^4 S_{mJy} c^2}{\nu^2 \vartheta_{min} \vartheta_{maj}} K$, using the major and minor axis $\vartheta_{min} \vartheta_{maj}$ of the restoring beam and the flux density $S_{mJy}$ at the observing frequency $\nu$. Similarly, the temperature limits of the spots can be determined using this relation by inserting the estimated size of the spots. We first performed image fitting to determine the size of the most prominent hotspot. Both using CASA and AIPS, the deconvolved component size is found to be indistinguishable from a point source. Thus, the spots are unresolved, and their size can be constrained to be a fraction of the beam size (see Method section 2). The temperature of the spots is then only a lower limit.

In order to determine the robustness of the structure seen in the continuum disk of W Hya, we also imaged each observing block separately. Although the beam is different for the three epochs, the structure is fully consistent. Similarly, the CO v=1, J=3-2 maps are consistent between observations.

### 2) Visibility fitting

Our fit to the stellar disk was performed using the visibility fitting code UVMULTIFIT *(32)*. The visibility fitting results are shown in Supplementary Figure 1 for the parameters described in the text. The error on the fits is dominated by the absolute flux calibration uncertainty. The disk temperature was determined using the

formula for the brightness temperature of an elliptical uniform disk *(33)*:

$$T_b = \frac{1.96 \cdot 10^4 S_{mJy} c^2}{\nu^2 \vartheta_{min} \vartheta_{maj}} K$$

The figure shows the azimuthally averaged visibilities after adjusting the phase center to the center of the stellar disk. It is clear that the disk fit accurately describes the largest scale structure (at the shortest baselines), but that there are significant compact components offset from the center of the disk. We have attempted to improve the fit by first including a single Gaussian component. We fit a disk with a fixed size of 51.0x56.5 mas (position angle 65.7 degrees), which provides a best fit flux density of 439 ± 44 mJy corresponding to a brightness temperature T=2380 ± 240 K. The Gaussian component is determined to have a size of $0.1^{+0.4}_{-0.1}$ mas and a flux density of 24.9 ± 0.3 mJy. The fit is presented in Supplementary Figure 2 and the residuals are shown in Supplementary Figure 3. The hotspot is thus clearly unresolved.

Because Supplementary Figure 3 shows several remaining features, we also performed a fit including several compact components. Included in Supplementary Figure 2, we introduce a model of the disk and a single compact unresolved spot in each stellar quadrant. We fit a disk with the same, fixed, size as above and find a flux density of 445 ± 45 mJy corresponding to a temperature T=2415 ± 242 K. The components are found with fluxe densities of 29.5 ± 0.3 mJy (SW), 3.7 ± 0.1 mJy (NW), -7.1 ± 0.3 mJy (NE), and -2.7 ± 0.3 mJy (SE). The residuals of this fit are shown in Supplementary Figure 4. While the detections of the weaker spots are all significant, the choice of a single component per quadrant is arbitrary and only serves to highlight that more than a single component is needed to describe the data. However, the degeneracies between models with an increasing number of free parameters are such that we cannot further constrain the surface structure. The actual residual image after subtracting the disk and single component (Supplementary Figure 2) is the best model for the remaining structure.

For the strongest hotspot, we can constrain the feature to be significantly smaller than the beam in both image and visibility fitting. The 1σ size limit is provided by $\sigma_s = \sqrt{2} \frac{\vartheta_{beam}}{SNR}$ *(34)*. Using the rms noise determined in the image, the strongest component has a signal-to-noise ratio SNR=59. For a beam of 17x24 mas, the one σ size limit is thus 0.4x0.6 mas. With 5 σ confidence we can thus determine the size to be <2x3 mas, providing a temperature of > $5.3 \times 10^4$ K.

### 3) CO J=3-2, v=1 modeling

To reproduce the observed vibrationally-excited rotational CO line at 342.648 GHz (v=1, J=3-2) *(35)*, we calculated radiative-transfer models consisting of a spherical symmetric circumstellar envelope extending radially from the measured $R_{*,338}$ out to the outer radius of the line emission region, $R_{out}$, at 2.4 $R_\star$ (4.8 au). The model is based on the one presented in *(27)*, but was modified to include three different layers, the outflowing gas, the inflowing gas, and the warm gas (T > 2400 K). The density and excitation temperature of CO in a given layer are assumed to be constant for simplicity. The gas mass is calculated using a abundance ratio of CO with respect to $H_2$ of $4 \times 10^{-4}$. A linear velocity profile, in which the velocity is proportional to the radius, is adopted and the module of the velocity decreases outwards for all layers. This choice is motivated by the velocity profiles found in wind-driving models *(36)*. The populations of the rotational levels were calculated using the Boltzmann distribution under the assumption of local thermodynamical equilibrium (LTE). The LTE assumption for the excitation of CO has been studied by *(37)*. The authors investigated the temperatures for the excitation of the vibrational levels and of the rotational levels within a given vibrational level. They find that the rotational-excitation temperature remains equal to the kinetic temperature of the gas down to densities of ~ $10^8$ cm$^{-3}$. This limit is more than two orders of magnitude below the densities we find. In their models, the vibrational-excitation temperature is not in LTE, however, for densities below, ~ $10^{12}$ cm$^{-3}$, which is larger than the densities we find by roughly a factor of ten. They find the vibrational-excitation temperature to be 20%-30% lower than the kinetic temperature. The temperature determined by our models using the v=1, J=3-2 transition of CO, is the rotational-excitation temperature of the CO molecules. If the vibrational-excitation temperature is indeed 20%-30% lower than the kinetic and the CO rotational-excitation temperature, we find that our models underpredict the mass in the extended atmosphere by at most about a factor of two.

We integrated the radiative transfer equation numerically for several lines-of-sight spanning the full radial extent of the envelope, from the center of the stellar disc to $R_{out}$. This was done at different frequencies so that the line was fully covered. Each line-of-sight was divided in (typically) a few hundred intervals over which the radial velocity component projected along the line-of-sight varied by only a small fraction. The optical depth of each interval at a given frequency was computed by considering the Doppler shift from the velocity component along the line-of-sight and the intrinsic line shape. The intrinsic line shape was considered to be a Gaussian function with standard deviation given by the quadratic sum of the standard deviation of two independent broadening terms, thermal motions and turbulence. The full-width at half maximum of the turbulence broadening term was considered to be 3 km s$^{-1}$ *(38)*, while the thermal broadening was calculated based on the temperature of the gas in a given layer.

When integrating the radiative transfer equation, the background intensity $I_o$ was considered to be the average of the continuum maps over each hemisphere for the lines-of-sight that intercepted the stellar disc, while it was set to zero for lines-of-sight that did not intercept the star. We convolved the intensity map produced by these calculations with a 17 mas beam before computing the flux densities that were compared to the observed values.

We calculated models to fit observed flux densities measured from two solid angles, towards the eastern and western hemispheres. The temperature, masses, and velocity range of the models for each hemisphere were constrained by fitting the emission towards and in an annulus around the star. In this way, the gas temperature and density could be uniquely determined. This approach assumes that models obtained for the eastern and western hemispheres contribute equally to the off-star emission. The observations support this choice, since there is no significant difference between off-star emission arising from the eastern and western hemispheres. We have tested a different approach in which the temperature and gas densities can still be uniquely determined, by considering the gas temperature to be equal between the two hemispheres. In this way, the contributions of the two regions to the off-star emission differ. We find that the derived temperatures and densities are different by only ~10% and ~5% with respect to the previous models.

We calculated models both with and without a layer of warm gas (T = 2900 K). This temperature is comparable to that found from infrared CO observations *(28)*. We find that including gas with a temperature higher than the background stellar continuum in the eastern hemisphere (T = 2250 K) produces emission and improves the fit (see Fig. 3). The temperature of this warm layer was set to the same value of the average stellar continuum temperature towards the western hemisphere (T = 2900 K), so that no emission would be produced towards this hemisphere. We constrain the possible existence of higher temperature molecular gas with a filling factor similar to that of the observed hotspots. The models without a layer of warm gas require temperatures and masses of cold gas that differ only slightly (Supplementary Table 1) from those found when the warm layer is included (Main text Table 1). The maximum outflow and inflow velocities derived for the cold gas are the same whether or not the warm gas layer is included. The warm layer accounts for roughly 10% of the off-star emission in our best-fit models.

We determined the best fits and estimated the uncertainties within the assumptions of our described model on the derived parameters by calculating the $\chi^2$ of the fit of the models to the observed spectra. We used spectra binned to 3 km/s resolution for these calculations to further improve the signal-to-noise. The expansion velocities of the outflow and inflow for each hemisphere were constrained by calculating the $\chi^2$ of the fit of models to the spectra observed towards the star for each given hemisphere. The temperatures and masses in each model component were constrained by calculating the $\chi^2$ of the fits to the spectra observed both towards the star for the given hemisphere and from around the star. For the eastern hemisphere, the best model without a layer of warm gas produces a reduced $\chi^2$ of 0.9 when compared to the spectrum towards the star (13 degrees of freedom) and of 2.1 when compared to the combination of spectra towards the star and from around the star (25 degrees of freedom). When a layer of warm gas is included, the reduced $\chi^2$ of the fits to the two datasets decrease respectively to 0.6 and 1.2 with 12 and 24 degrees of freedom. For the West hemisphere, the models with and without the warm layer produce a reduced $\chi^2$ of 1.1 and 1.9 respectively when compared to the spectrum towards the star (11 degrees of freedom) as well as when compared to the spectra towards the star and from around the star (23 degrees of freedom) together.

Thus, although the fits are, in most cases, not statistically distinguishable when fitting the entire spectrum, the models including a warm layer produce improved fits. This difference is particularly noticeable around the stellar velocity in the eastern hemisphere. Here several channels of emission appear that cannot be readily produced without a warm layer even if the effect on the global $\chi^2$ calculation is minimal. Based on the channel noise of 3.7 mJy in the spectrum, we calculate that the integrated emission in those channels constitutes a 4.2σ detection. To determine the robustness of this detection we perform similar calculations for other features in the spectra that span more than four channels, but none of these exceed a 2σ detection threshold. As noted above, the temperature of our best fit warm layer is also consistent with similar warm layers found from infrared CO observations *(27)* and hence favor the models including the warm layer.

Our best models have maximum off-star optical depths of about 1, and, hence, the circumstellar envelopes are not very optically thick in this vibrationally-excited CO line. The lines-of-sight towards the star show lower optical depths with a maximum of about 0.3.

**4) Adjustment of the stellar velocity of W Hya**

Based on CO molecular transitions in the ground vibrational state, the stellar velocity of W Hya with respect to the local standard of rest, was found to be $V_{*,lsr}$ = 40.4 km s$^{-1}$ *(18)*. However, it was previously noted that the higher excitation transitions, originating closer to the star, were better reproduced using a slightly lower velocity $V_{*,lsr}$ = 39.6 km s$^{-1}$ *(18)*. Additionally, a large number of maser lines show narrow components at lower velocity *(39)*. Using the emission from the CO v=1, J=3-2, line presented here, we find that it peaks around V = 39.2 ± 0.2 km s$^{-1}$. We have investigated the effects that could produce this apparent shift in velocity. We can rule out the effect of a high line optical depth in the stellar limb, as it is not consistent with the absorption spectra. Alternatively, a lower gas density behind the star or a preferred blue-shifted motion of the atmosphere could cause the emission to peak blue-shifted from the stellar velocity. However, it would be highly unlikely to see this effect along the entire stellar limb. We thus suggest that the true stellar velocity is $V_{*,lsr}$ = 39.2 ±

0.2 km s$^{-1}$. This also fits with the previous observation that the higher excited lines, and several masers, peak at a velocity below 40 km s$^{-1}$.

### 5) An unidentified line in the hot stellar atmosphere

Across the observed spectral windows, we detect one line in emission in front of the stellar continuum, with a signal-to-noise ratio at the peak of 6.5 (Supplementary Figures 5, 6). This emission peaks in two positions, one towards the south-west, hot part, and one towards the north-east, cooler part of the disk (Figs. 1, Supplementary Figure 5).

Assuming the corrected stellar velocity $V_{*,lsr} = 39.2$ km s$^{-1}$, we tentatively identify this emission as the rotational transition $J_{Ka,Kc} = 27_{3,25} - 28_{0,28}$ in the vibrationally excited SO$_2$($v_2$=2) bending mode. This transition has a tabulated rest frequency of 345.01713 GHz, with a large uncertainty of 5.702 MHz (*40*), whereas the centroid frequency of the observed spectrum is 345.016 GHz. The transition has an upper-level energy $E_{up}/k$ of 1832 K and an Einstein-*A* coefficient of 2.2x10$^{-6}$ s$^{-1}$ (*41*). We use the ExoMol database for the Einstein coefficients and energy level energy. The ExoMol line list is generated using the DVR3D code, which computes rovibrational transition frequencies and Einstein coefficients from a potential energy and dipole moment surface, instead of using a phenomenological fitting Hamiltonian. For rotational transitions within one vibrational state, this program is expected to have a relatively large error on the transition frequencies, but will yield the very accurate Einstein coefficients.

Two additional transitions in the SO$_2$($v_2$=2) state were observed but not detected in this data set: $J_{Ka,Kc} = 38_{5,33} - 38_{4,34}$ at 345.38558 GHz (with a 7.551 MHz uncertainty) and $J_{Ka,Kc} = 43_{5,39} - 44_{2,42}$ at 345.56536 GHz (with a 9.719 MHz uncertainty) with respective upper-level energies $E_{up}/k$ of 2209 K and 2369 K and Einstein-*A* coefficients of 4.0x10$^{-4}$ s$^{-1}$ and 5.8x10$^{-6}$ s$^{-1}$. The upper-level energy of the detected line and the fact that the line is in emission in front of the stellar disk could imply excitation in an environment hotter than the star. If the line is indeed SO$_2$($v_2$=2), this is supported by the narrow line-width (7.3 ± 0.6 km s$^{-1}$, Supplementary Figure 6), and by the velocity coincidence of this emission with the warm CO layer. However, since Einstein-*A* coefficients are proportional to the rate of stimulated emission, the low Einstein-$A$ coefficient for the detected transition could imply maser action. This would explain the non-detections of either of the other two SO$_2$($v_2$=2) lines even though their upper-level energies correspond to temperatures reached across the stellar disk. However, the rather large linewidth is not consistent with strong maser action.

We also detect the $J_{Ka,Kc}$=23$_{3,21}$ – 23$_{2,22}$ transition in the SO$_2$($v_2$=1) state at 342.436 GHz, with an upper-level energy of 1021 K, which supports the excitation and detectability of vibrationally excited SO$_2$. This transition is in absorption in front of the stellar disk and in emission around it.

If our identification is correct, this is the first detection of emission from the excited SO$_2$($v_2$=2) state towards and AGB star. However, considering the non-detection of other SO$_2$($v_2$=2) lines and the large uncertainties on the transition frequencies, we also cannot rule out that the emission originates from a different, unidentified, molecular or atomic transition. Nonetheless, the fact that the line is seen in emission even against the hottest part of the star supports the existence of hot gas.

**Data availability:** The data that support the findings of this study are available though the ALMA archive (https://almascience.eso.org/alma-data/archive) with project code: ADS/JAO.ALMA#2015.1.01446.S

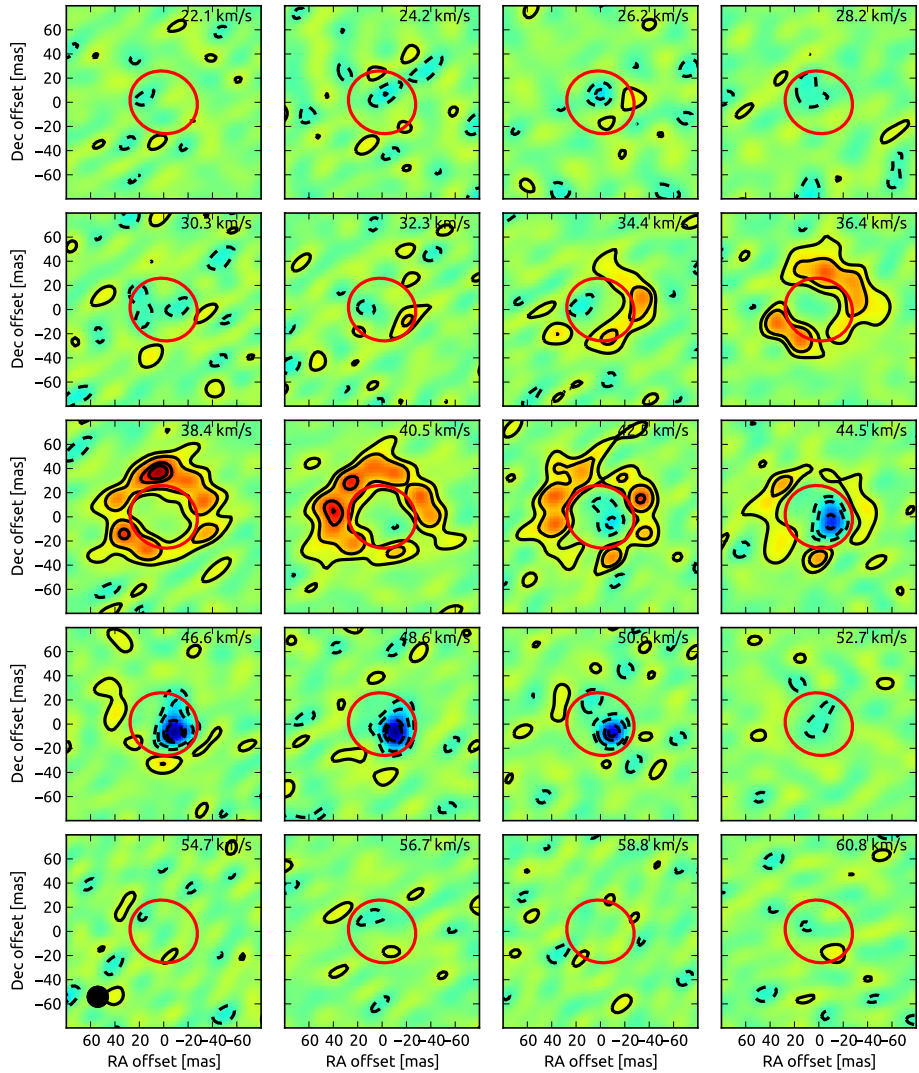

**Supplementary Figure 1.** Channel maps of the vibrationally excited pure-rotational transition of CO J=3-2, v=1 at 342.646 GHz around W Hya. As in Fig.1, the images were made using uniform weighting and restored with a circular 17 mas beam. Contours are drawn at -8, -6, -4, -2, 2, 4, 6, and 8 times the rms channel noise, which is 2.6 mJy Beam$^{-1}$. The velocity resolution for these images was reduced to ~2 km s$^{-1}$. The ellipse in each panel indicates the stellar disk at 338 GHz. The maps are dominated by red-shifted absorption, and the emission peaks around the stellar velocity, determined to be $V_{*,lsr}$ = 39.2 km s$^{-1}$.

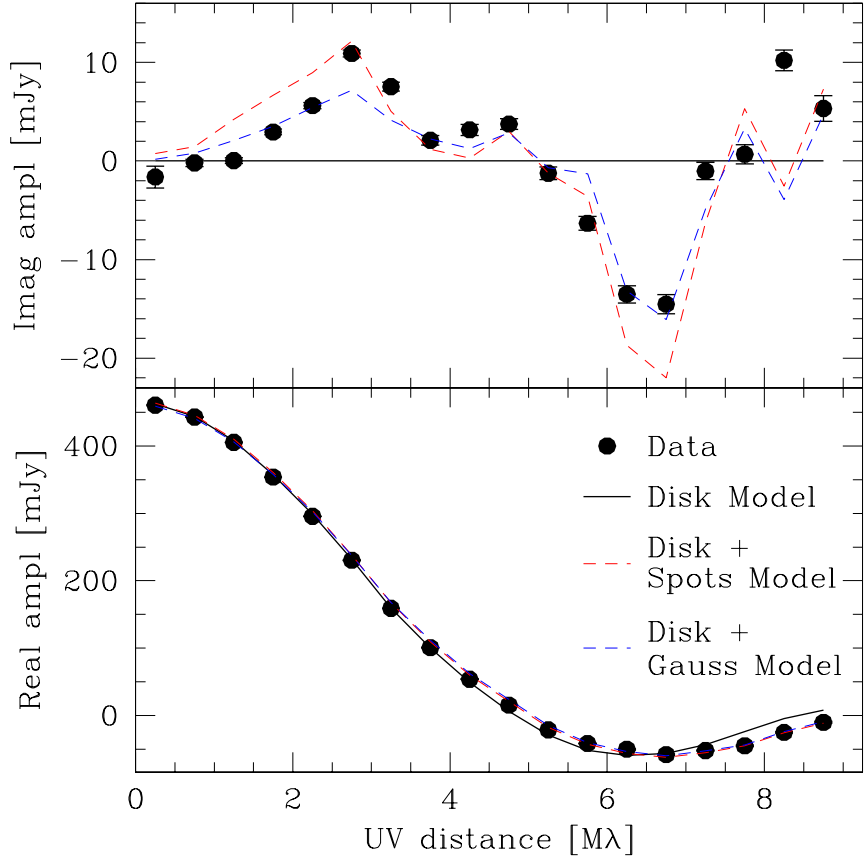

***Supplementary Figure 2***. Azimuthally averaged Real (bottom) and Imaginary (top) amplitudes of the source visibilities against baseline length. In blue, we include the model of a disk with a single Gaussian component that is consistent with a point source. In red, we also include the best-fit disk model with 4 additional compact components. The parameters for both models are described in the text. The residuals of the disk + single component fit are shown in Supplementary Figure 3, and the residuals of the disk + multiple component fit are shown in Supplementary Figure 4. The error bars on the data points are indicated and in most cases are smaller than the symbol size. The deviations from a uniform disk are obvious both at long baselines in the Real amplitudes and in particular in the Imaginary amplitudes. As indicated by the fit including the compact spots, the imaginary amplitudes show a pattern consistent with off-center hotspots across much of the baseline range. Specifically the residual maps in Supplementary Figures 3 and 4 show the improvements when including more than a single component.

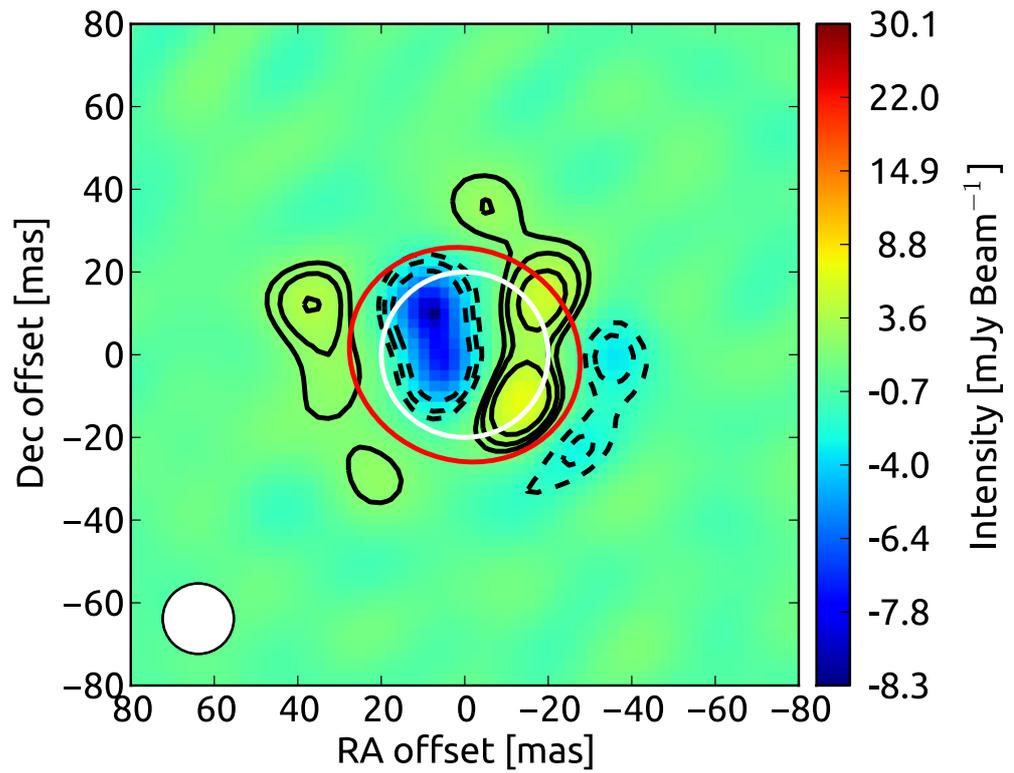

***Supplementary Figure 3.*** *The residual map after subtracting a uniform elliptical disk and a single Gaussian component from the continuum image of W Hya at 338 GHz. The size of the Gaussian component is fit to be $0.1^{+0.4}_{-0.1}$ mas and is thus consistent with an unresolved component. Contours are drawn at -9, -6, -4, 4, 6, 9 and 12 times the image rms noise of 0.45 mJy Beam$^{-1}$. Compared to Fig.2 of the main text, the residuals have decreased. Significant (>6σ) deviations from a uniform disk are still visible.*

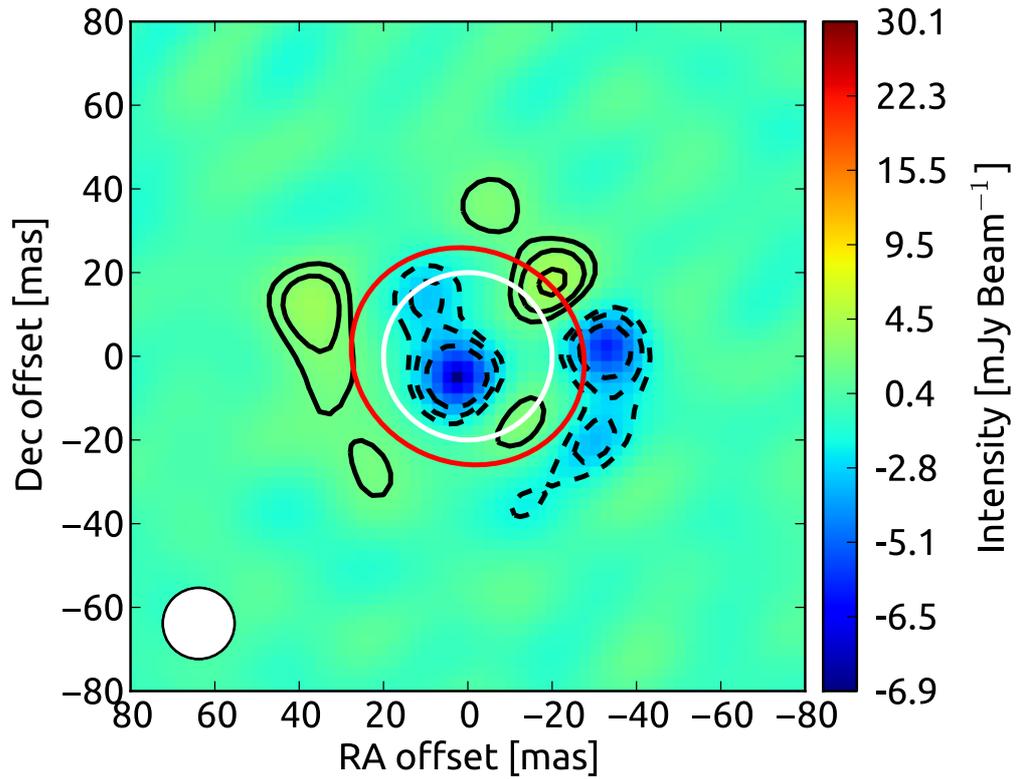

***Supplementary Figure 4.*** *The residual map after subtracting a uniform elliptical disk and four compact components from the continuum image of W Hya at 338 GHz. Contours are drawn at -9, -6, -4, 4, 6, 9 and 12 times the image rms noise of 0.45 mJy Beam$^{-1}$. Compared to Supplementary Figure 3, the residuals have decreased by 30%, highlighting the existence of more than one component. However, further deviations from a uniform disk are still obvious.*

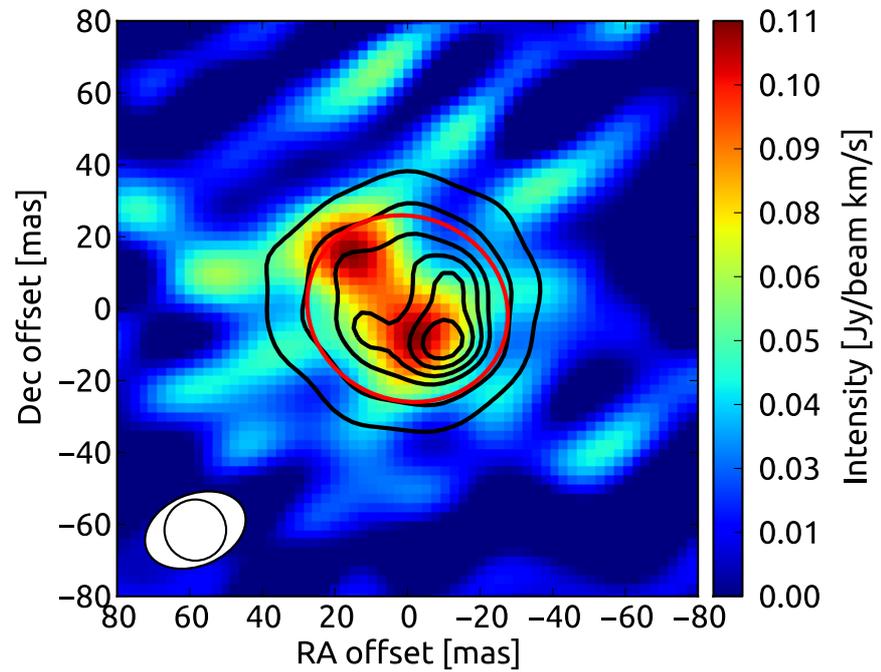

***Supplementary Figure 5***. The background image is the integrated intensity of the unidentified line that is seen in emission against the warm stellar continuum (shown in black contours). The red ellipse indicates our fit to the stellar disk. The line is tentatively identified as the rotational transition $J_{Ka,Kc} = 27_{3,25} - 28_{0,28}$ in the vibrationally excited $SO_2(v_2=2)$ bending mode. It appears to originate from the hot molecular gas layer. The dirty beam size and the adopted restoring beam are shown in the bottom left.

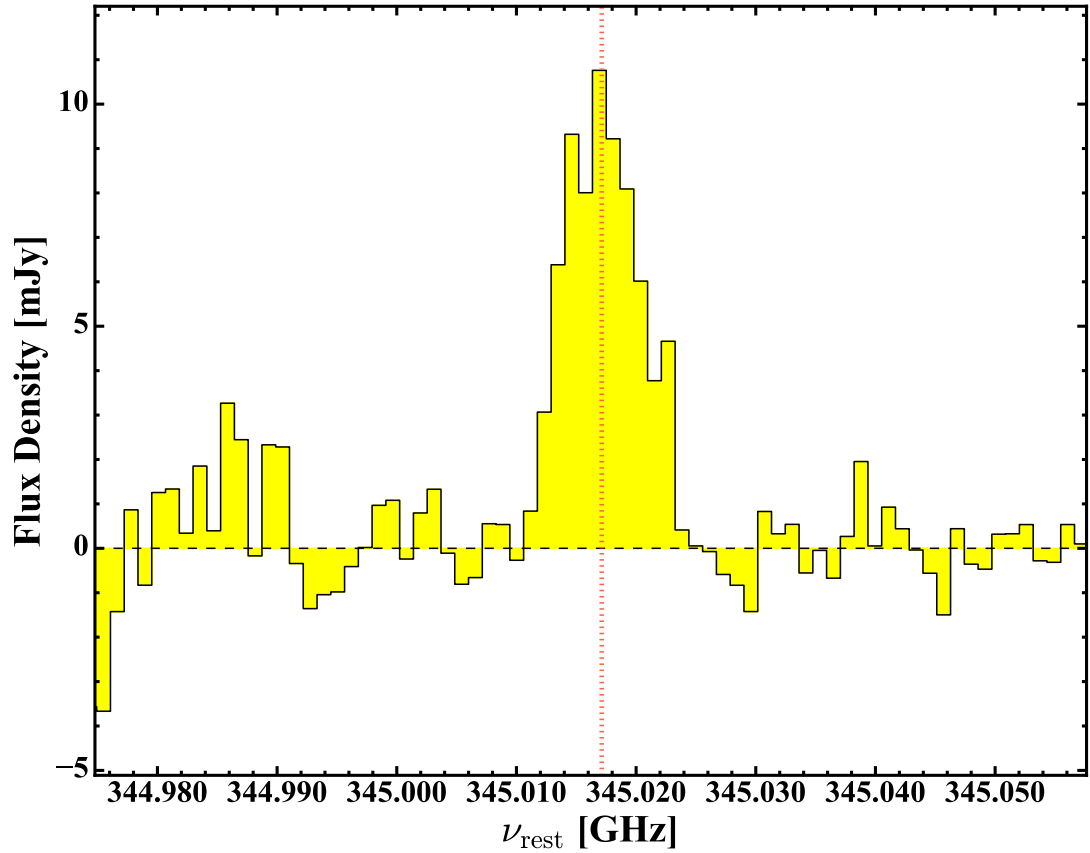

***Supplementary Figure 6***. *The rest frame spectrum of the unidentified emission line extracted against the stellar continuum. The spectrum is corrected for a stellar velocity of $V_{*,lsr}$ = 39.2 km s$^{-1}$. The vertical line indicates the rest frequency of the transition $J_{Ka,Kc} = 27_{3,25} - 28_{0,28}$ in the vibrationally excited $SO_2(v_2=2)$ bending mode of 345.01713 GHz*

| Hemisphere | Component | Temperature [K] | Mass [M$_\odot$] | Velocity Range [km s$^{-1}$] |
|---|---|---|---|---|
| **East** | **Infall** | 1000 ± 100 | (3.5 ± 1.2) x10$^{-5}$ | (-21 ± 3) to (-2 ± 1) |
| | **Outflow** | 1000 ± 100 | (1.9 ± 0.6) x10$^{-5}$ | (+3 ± 1) to (+20 ± 3) |
| **West** | **Infall** | 850 ± 50 | (7.0 ± 1.0) x10$^{-5}$ | (-13 ± 1) to (-2 ± 1) |
| | **Outflow** | 850 ± 50 | (5.0 ± 1.0) x10$^{-6}$ | (+13 ± 2) to (+20 ± 3) |

***Supplementary Table. 1.*** *Mass, average temperature, and velocity range for the molecular gas components that best match the observed spectra of the CO v=1, J=3-2 transition shown in Fig.3 without including a warm layer. The velocities indicate outflow (positive) and infall (negative) with respect to V$_{*,lsr}$ = 39.2 km s$^{-1}$.*